\documentclass[iop]{emulateapj}

\slugcomment{Accepted for publication in the Astrophysical Journal Letters}

\shorttitle{A tidally disrupted dwarf galaxy in the Hydra I Cluster}
\shortauthors{A. Koch et al.}

\begin{document}

\title{Threshing in Action -- The tidal disruption of a dwarf galaxy by the Hydra I Cluster\altaffilmark{$\dagger$}}	% emulateapj

\author{
Andreas Koch\altaffilmark{1},  
Andreas Burkert\altaffilmark{2,3}, 
R. Michael Rich\altaffilmark{4}, 
Michelle L.M. Collins\altaffilmark{5}, 
Christine S. Black\altaffilmark{4},\\  
Michael Hilker\altaffilmark{6}, 
\& 
Andrew J. Benson\altaffilmark{7}
}
\altaffiltext{$\dagger$}{Based on observations carried out at the European Southern Observatory under proposal number 65.N-0459(A).}
\altaffiltext{1}{Zentrum f\"ur Astronomie der Universit\"at Heidelberg,  Landessternwarte, K\"onigstuhl 12, 69117 Heidelberg, Germany}
\altaffiltext{2}{Universit\"atssternwarte der Ludwig-Maximilians Universit\"at, Scheinerstr. 1, 81679 M\"unchen, Germany}
\altaffiltext{3}{Max-Planck-Fellow; Max-Planck Institute for Extraterrestrial Physics, Giessenbachstrasse, D-85748 Garching, Germany}
\altaffiltext{4}{University of California Los Angeles, Department of Physics \& Astronomy, Los Angeles, CA, USA}
\altaffiltext{5}{Max-Planck-Institut f\"ur Astronomie, K\"onigstuhl 17, 69117 Heidelberg, Germany}
\altaffiltext{6}{European Southern Observatory, Karl-Schwarzschild-Strasse 2, 85748 Garching, Germany}
\altaffiltext{7}{Department of Astronomy, Caltech, Pasadena, CA, USA}
\email{akoch@lsw.uni-heidelberg.de}
\begin{abstract}
We report on the discovery of strong tidal features around a dwarf spheroidal galaxy 
in the Hydra I galaxy cluster, indicating its ongoing tidal disruption.
This very low surface brightness object, HCC-087, was originally classified as an early-type dwarf in the Hydra Cluster Catalogue (HCC), but 
our re-analysis of the ESO-VLT/FORS images of the HCC unearthed a clear indication of an S-shaped morphology and a large spatial extent. 
Its shape, luminosity (M$_V=-11.6$ mag),  and physical size (at a  half-light radius of 3.1 kpc 
and a full length of $\sim$5.9 kpc) are comparable to the recently discovered NGC~4449B and the Sagittarius dwarf spheroidal, 
all of which are undergoing clear tidal disruption.  
Aided by N-body simulations we argue that  HCC-087 is currently at its first apocenter, at 150 kpc, around the cluster center and that it is being tidally disrupted by the 
galaxy cluster's potential itself.  
An interaction with the near-by (50 kpc) S0 cluster galaxy HCC-005, at M$_{\ast}\sim3\times10^{10}$ M$_{\odot}$ is rather unlikely, 
as this constellation requires a significant amount of dynamical friction and thus low relative velocities. 
The S-shaped morphology and large spatial extent of the satellite
would, however, also appear if HCC-087 would orbit the cluster center. 
These features appear to be characteristic properties of satellites that are seen in the process of being tidally disrupted, independent of the environment of the destruction. 
An important finding of our simulations is an orientation of the tidal tails perpendicular to the orbit.
\end{abstract}
\section{Introduction}
Tidal sculpting and the disruption of satellite galaxies  is a prevalent source for the hierarchical build-up of larger galaxies' halos (e.g., Searle \& Zinn 1978).  
The action of tides in the Local Group (LG) is in fact observed through the accretion of the Sagittarius galaxy (Ibata et al. 1994) and Andromeda's giant stellar stream (Ibata et al. 2001).
Over the past years, tidal features manifested as low surface brightness structures surrounding near-by galaxies in the Local Volume (e.g., Mart{\'{\i}}nez-Delgado 2010), and out to higher redshifts (Forbes et al. 2003; Sasaki et al. 2007; 
Arnaboldi et al. 2012), lending strong support to the 
hierarchical scenario of galaxy formation (e.g., Boylan-Kolchin et al. 2010).

The recent discovery of a disrupted satellite to the Magellanic irregular starburst galaxy NGC 4449 (Rich et al. 2012; Mart{\'{\i}}nez-Delgado et al. 2012) provides the first evidence that such accretion events  already take place on 
remarkably small scales. This satellite exhibits an S-shaped morphology that is observed and predicted for tidally disrupting star clusters and  galaxies (e.g., Capuzzo-Dolcetta et al. 2005; Pe\~narrubia et al. 2009). 
From a modeling point of view, tidal effects induce 
significant variations  of a systems'  characteristic radius (e.g., the half-light radius, $r_h$). 
For instance, it has been noted that the brighter satellite galaxies to M31 are more extended than their Milky Way (MW) counterparts -- possibly induced by its massive disk (Hammer et al. 2007; Pe\~narrubia 2010; cf. Richardson et al. 2011). 

In this context, scaling relations for ``dynamically hot'' systems, i.e., those supported by random motions, provide
an important tool for studying the underlying physics of structure formation
 (e.g., Gilmore et al. 2007; Dabringhausen et al. 2008; Misgeld \& Hilker 2011; Brasseur et al. 2011). 
 Correlations between, say, luminosity and radius  were found over a wide mass range, from giant ellipticals down to the dwarf galaxy regime. 
Moreover,  star clusters and ultracompact dwarfs (UCDs), i.e., low-luminosity, spatially concentrated objects, are clearly differentiated in this parameter space.  
A characterisation of {\em extended}, low-surface brightness objects, however,  is still incomplete\footnote{It should be noted, however, that such scaling relations often hold a strong detection bias against extended, low surface brightness targets. For instance, the  surface brightness detection limit 
of the Sloan Digital Sky Survey lies at $\mu_V\sim 30$ mag\,arcsec$^{-2}$ .},  
while those objects hold important clues to the environmental effects  governing galaxy evolution. 

Of further interest is the evolution of the galaxies themselves as they undergo ÒthreshingÓ in the potential of their hosts: for instance, this way a transformation of 
a late-type spiral with a nuclear  star cluster (Sc,N) 
into nucleated dwarfs can occur, until they  
may further morph into UCDs (e.g., Goerdt et al. 2008), 
which, in turn, may play an important role in the build-up of massive globular cluster systems. Clearly, this complex interplay emphasizes the importance of studying tidally disrupting systems for a deeper understanding of structure formation. Unfortunately, the time-scales for such interactions are yet poorly known and basically framed only by simulations (e.g., Pe\~narrubia et al. 2008, 2009). 

Here we report on the discovery of an extended, early-type galaxy in the Hydra I Cluster ($\equiv$Abell 1060) that shows clear evidence of tidal disruption.  
Although its shape and luminosity parameters have been listed in the Hydra Cluster Catalogue (HCC; Misgeld et al 2008, hereafter M08), its distorted appearance had so far eluded detection.
No other such remarkable object has been found in this cluster, making it a unique object in the Local Volume, together with NGC~4449B. 
Hydra~I is an evolved and dynamically relaxed, yet poor galaxy cluster, dominated by early-type galaxies, 
at a distance of 47.2 Mpc (Misgeld et al. 2011). 
\section{Data}
The images presented herein were 
first  published in Mieske et al. (2005) and M08. 
In brief, the exposures were taken with VLT/FORS1 at ESO/Paranal in April 2000.
``Field 6'', which also contains HCC-087, was observed in the I-band (9$\times$330 s)  on April 5, 2000 
and on April 6, 2000  in the V-band (3$\times$480 s). 
The conditions were mostly photometric, at an average seeing of 0.6$\arcsec$ in both V and I.
For details on the photometric  calibrations of those data we refer the reader to  Mieske et al. (2005) and M08. 
Figures~1 and 2 show the relevant image of the field surrounding HCC-087. 
\begin{figure}[htb]
\begin{center}
\includegraphics[angle=0,width=1\hsize]{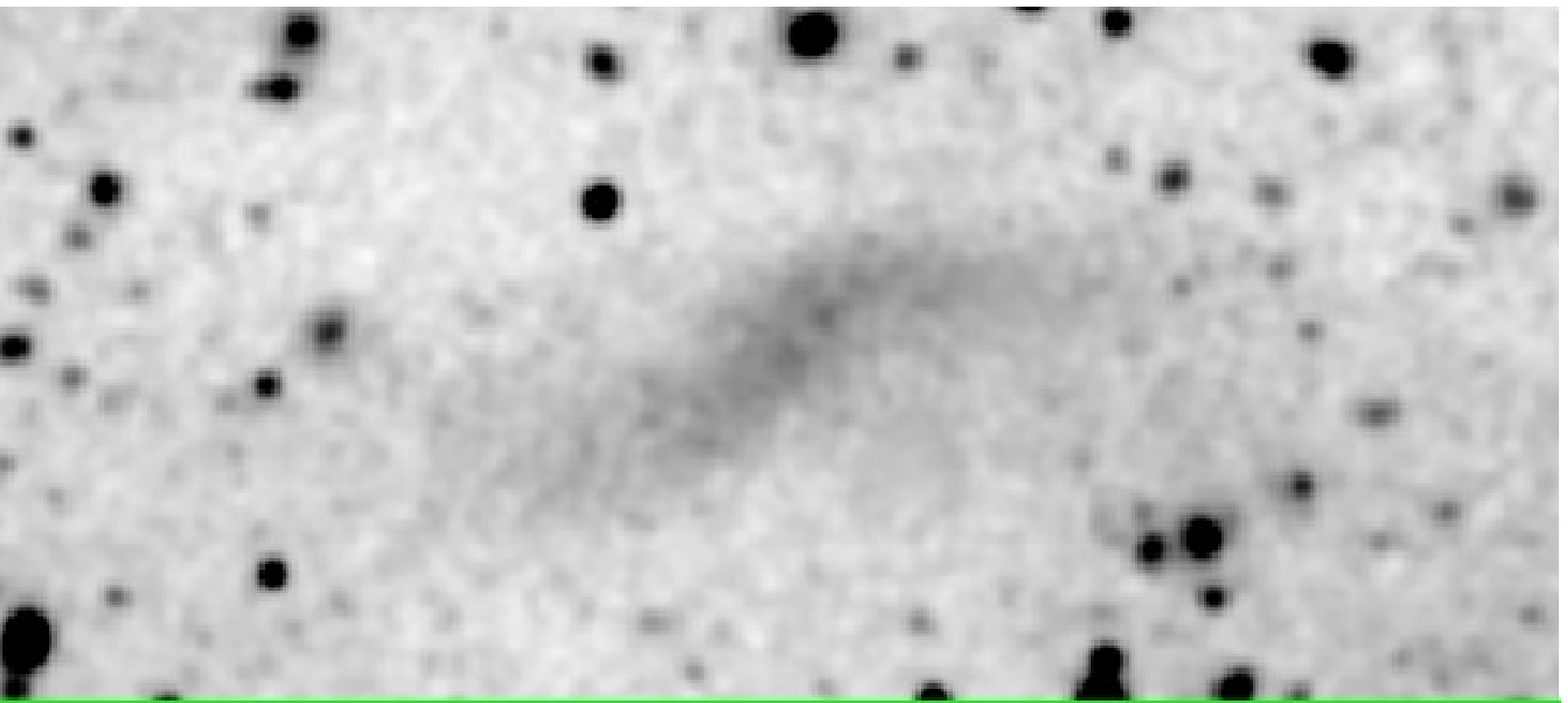}
\includegraphics[angle=0,width=1\hsize]{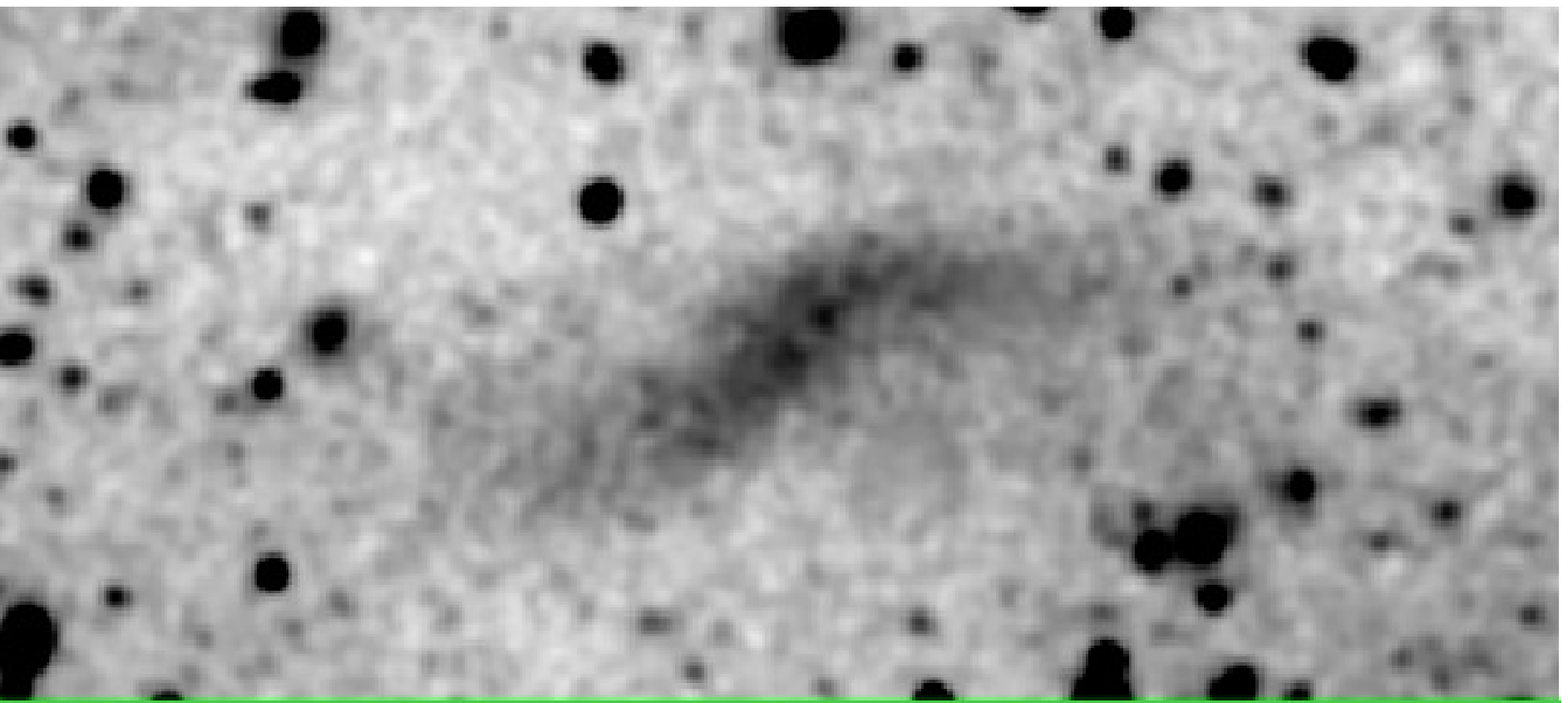}
\includegraphics[angle=0,width=1\hsize]{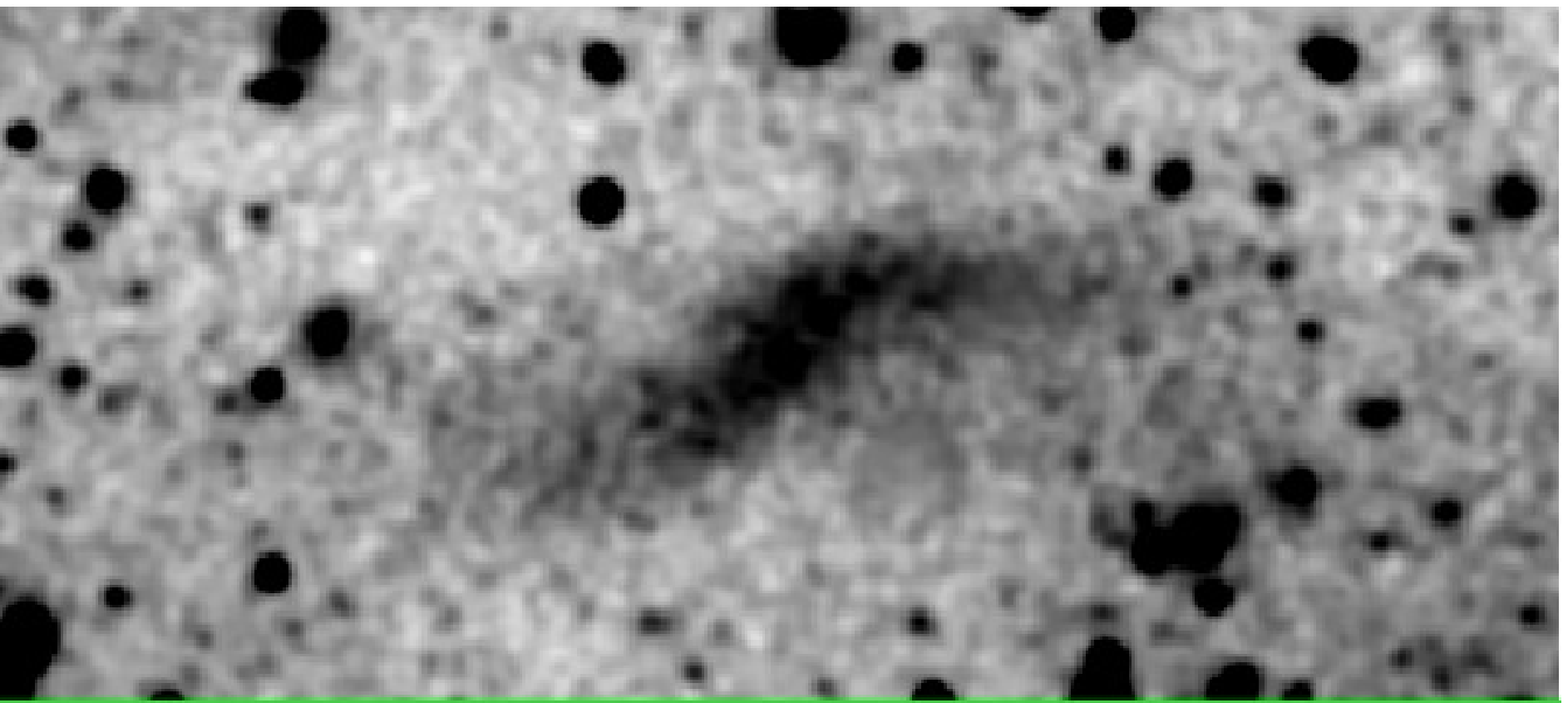}
\end{center}
\caption{Co-added V and I images of  HCC-087 using different stretches to enhance its morphology. These images cover $1.6\arcmin\times0.7\arcmin$  ($\sim22\times10$ kpc) 
and were smoothed with a Gaussian kernel.  Some bright stars close to the galaxy were PSF-subtracted from the images.}
\end{figure}
\begin{figure}[htb]
\begin{center}
\includegraphics[angle=0,width=1\hsize]{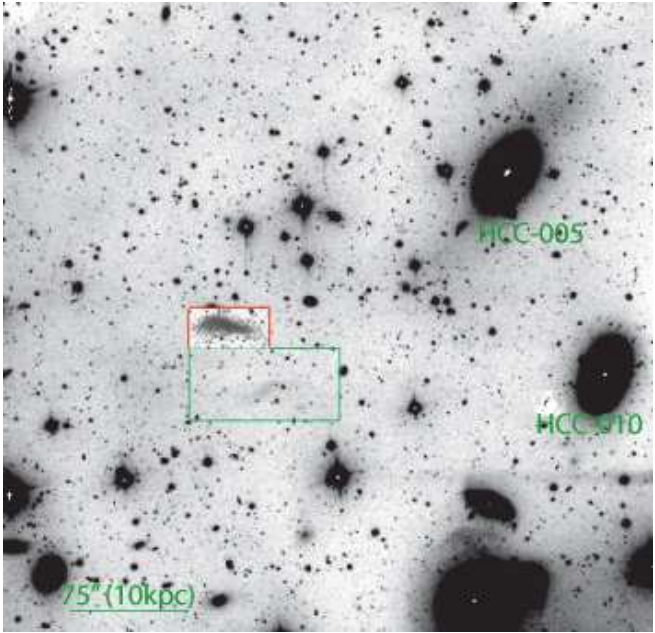}
\end{center}
\caption{ $6.8\arcmin\times6.8\arcmin$ ($\sim$93 kpc $\times$ 93 kpc) V-band image of the region around HCC-087 with the section from Fig.~1 highlighted as a green box.  The red-boxed inset shows NGC 4449B from Rich et al. (2012), scaled to the physical size at the distance of HCC-087. 
The scale bar indicates 
75$\arcsec$ (10 kpc at the distance of Hydra I). North is up, East to the left. We also label HCC-005 and HCC-010, as discussed in the text. 
Not visible in this field are Hydra's main galaxies NGC~3309, 3311, and 3312.}
\end{figure}

M08 performed (elliptical) isophote fitting on locally sky-subtracted images. Coupled with visual inspection, HCC-087 was originally classified as an early-type dwarf galaxy and 
its photometric and morphological parameters are $\mu_0$=$26.3$ mag\,arcsec$^{-2}$ and $R_{\rm eff}$=13.7$\arcsec$ (Table~1). 
At an estimated M$_V$ of $-11.6$ mag, it is comparable to the luminous MW satellite Leo~I (e.g., Koch 2009). 
\begin{center}
\begin{deluxetable}{ccc}
\tabletypesize{\scriptsize}
\tablecaption{Properties of HCC-087 and NGC~4449B}
\tablewidth{0pt}
\tablehead{ & \colhead{HCC-087}  & \colhead{NGC~4449B} \\
\raisebox{1.5ex}[-1.5ex]{Parameter}& \colhead{M08}  &\colhead{Rich et al. (2012)}}
\startdata
$\alpha$, $\delta$ (J2000.0) &  10:36:39.0, $-$27:21:25.5 & 12:28:45, 43:57:44 \\
V, V$-$I [mag]               & 21.75, 1.07 & 14.88,  0.48 (g$-$r)\\
M$_V$ [mag]                  & $-$11.62$\pm$0.09 & $-13.03$$\pm$0.10\\
$\mu_0$ [mag\,arcsec$^{-2}]$         &    26.28$\pm$0.05 & 25.5$\pm$0.05 \\
$\Sigma_0$ [L$_{\odot}$\,pc$^{-2}]$  &     0.89$\pm$0.04  & 0.44$\pm$0.02 \\
r$_{\rm h}$ [arcsec]               &    13.65$\pm$4.62  & 147$\pm$14\\
r$_{\rm h}$ [kpc]\tablenotemark{a} &      3.1$\pm$1.1  & 2.72$\pm$0.16
\enddata
\tablenotetext{a}{Adopting a distance modulus of 33.37 mag for HCC-087 (Misgeld et al. 2011) and of 
27.91 for NGC 4449B (Annibali et al. 2008).}
\end{deluxetable}
\end{center}
%
%%%%%%%%%%%%%%%%%%%%%%%%%%%
%
%
Its extraordinarily large radius of 3.1 kpc (note that the average radius of the remainder of the HCC dwarfs lies at 650 pc) 
and its similarity with the strongly tidally affected NGC 4449B (Rich et al. 2012) prompted us to revisit those images, whereupon the tidal nature of HCC-087 became clear. 
We emphasize that the clearly distorted shape is visible on both the V- and I-band images, ruling out an instrumental artefact. 
Unfortunately, no radial velocity measurement is available for this galaxy, nor is that straightforward to obtain, given 
the object's faintness. We note, however, that there is no current evidence of any fore- or background structures within $\sim$50 Mpc relative to Hydra  (e.g., Richter 1989; M08). 
\begin{figure}[t!]
\begin{center}
\includegraphics[angle=0,width=1\hsize]{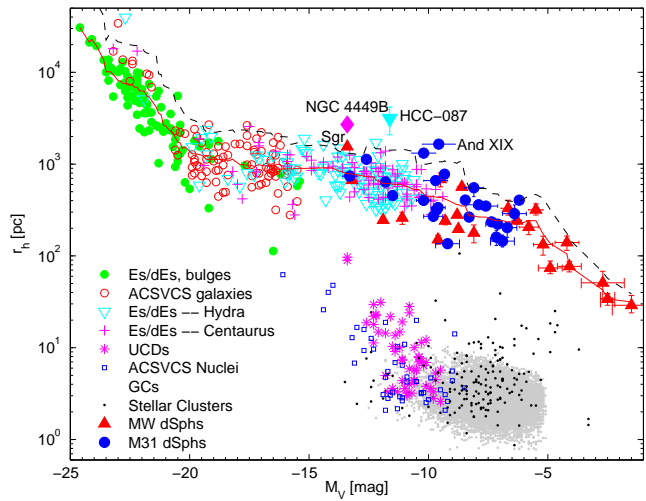}
\end{center}
\caption{Radius-magnitude plot, adopting data and classifications from Misgeld \& Hilker (2011). Highlighted are the tidally disrupted HCC-087, NGC 4449B (Rich et al. 2012), and other remarkably extended galaxies. The solid line indicates the 
median relation defined by the upper branch in the $r_h$-$M_V$-plane and the dashed line delineates the 2$\sigma$-upper limit to this trend. Note that the 
decline at the faint end is purely a detection bias, reflecting the surface brightness limit.}
\end{figure}
\section{Environment and tidal disruption}
\subsection{Morphology}
HCC-087 shows a clear S-shaped morphology, as reproduced in models of tidally disrupting dwarf satellites; the tails are aligned East-to-West. 

The effective radius of HCC-087 from an exponential profile fit is 3.1$\pm$1.1 kpc (M08). 
This is about the same value
found for the dwarf galaxy NGC 4449B (Rich et al. 2012)  that is currently undergoing strong tidal disruption 
on its first encounter with the luminous Magellanic galaxy NGC 4449 (M$_B = -18.2$ mag). 
Moreover, this value exceeds that of the 
(bound region of the) 
Sagittarius  dwarf by at least one kpc (Majewski et al. 2003; Walker et al. 2009), 
rendering HCC-087 and NGC 4449B the largest  such 
measurements for dwarf galaxies (Fig.~3). 
Furthermore, we estimate that the main body of HCC-087 is approximately $24.5\arcsec\times 8\arcsec$, or $5.6\times 1.8$ kpc,  in extent. 

In fact, both HCC-087 and NGC~4449B exceed the average radii of cluster galaxies in the Local Volume,  as sampled 
by the Hydra I and Centaurus cluster members and the M31 dwarf spheroidal companions,  by more than 2$\sigma$ (accounting for measurement errors {\em and} the scatter in the  $r_h$-$M_V$-plot). This is 
emphasized by the solid and dashed lines in Fig.~3,  which show the running median and 2$\sigma$ limits of these galaxies.

While the significantly greater extent of a number of brighter M31 dwarfs compared to
those of the MW has been noted by McConnachie \& Irwin (2006), we label in Fig.~3 the extreme M31 structural outlier, And~XIX (McConnachie et al. 2008). 
Although it has been suggested that tidal interactions with the disk of M31 are a possible cause for the larger extent (e.g., Collins et al. 2011), an assessment of the tidal nature of this satellite has to await 
measurements of its kinematics. 
\subsection{HCC-087 within the Hydra Cluster}
Lacking a redshift measurement and in the light of the richness of the cluster field surrounding HCC-087 it is difficult to unambiguously identify an obvious parent galaxy that is host to interactions with 
this dwarf galaxy. This is aggravated by the fact that 
the tails are seen in projection, while, arguing about  the entire Hydra~I cluster, we are
in fact dealing with a three-dimensional structure. 

We note a clustering of several elliptical galaxies NW to SW of HCC-087 and at projected distances of 35--50 kpc in Fig.~2, 
but none of the Hydra censuses lists any member systems closer than 30 kpc (Richter 1987; Christlein \& Zabludoff 2003; HCC).  
It could be attractive to associate the brightest of the ellipticals in the SW corner of Fig.~2 with  HCC-087 as this object is surrounded by a pronounced shell, indicative of tidal interactions (e.g.,  Mori \& Rich 2008; McConnachie et al. 2009; Rich et al. 2012). 
However, this object lies at a redshift of $z=0.03$   (Smith et al. 2004) and, together with the galaxy 1$\arcmin$ to its North, is a confirmed non-member of the Hydra cluster and thus can be ruled out as interacting partner. 

HCC-087 lies at a {\em projected} 140 kpc from the cluster's primary galaxies, NGC 3309 (E3) and NGC 3311 (S0), and at 170 kpc from the {\em Sab} LINER galaxy NGC 3312 (not pictured in Fig.~2). 
This 
is comparable to  the location of the Hercules dwarf spheroidal satellite within the  MW halo (although the LG clearly provides a different environment), which has been suggested to be a stellar, tidal stream in the process of formation (Martin \& Jin 2010). This provides the same order of magnitude as is seen 
in interacting dwarf-giant galaxies at higher redshifts (Sasaki et al. 2007). 

This is in contrast to NGC 4449B lying at 9 kpc from  NGC 4449 (albeit lower-mass than the luminous Hydra galaxies) and being on its first infall path towards its disruptive host.  
Rich et al. (2012) suggested from the appearance of this dwarf's tidal tails that it must be on a highly eccentric orbit and is currently 5--10 crossing times (or 200 Myr) 
past perigalacticon; this is also the 
time scale needed for transversing the (projected) distance between the partner and the dwarf galaxy under reasonable assumptions for its orbital motion.

There are two other HCC early-type galaxies, $\sim$50 kpc in projection to the W and NW of the dwarf -- HCC-010  and HCC-005, respectively, with stellar masses of a few times 
10$^{10}$M$_{\odot}$ (Misgeld \& Hilker 2011).  
Based on the orientation of the tidal tails and aided by our N-body simulations (Sect.~3.3), we will continue by considering both the cluster's central potential and HCC-005 as possible hosts to 
the tidal disruption of HCC-087, as we elaborate in the following.
\subsection{Simulations of tidal disruption}
A clearly boxy shape of the envelope (third panel of Fig.~1) 
is reminiscent of simulations of tidally disrupting galaxies 
(e.g., Bekki et al. 2001;  Pe\~narrubia et al. 2008), all of which bear closest resemblance with HCC-087 
near pericenter after an interaction time scale of the order of  500 Myr. 
The lack of an obvious, {\em very near-by} parent for HCC-087, however, would be  worrisome if it was at pericenter, even more so considering the ephemerality 
 of the tidal arm features in those simulations: the S-shaped morphology of threshed galaxies in the simulations of, say, Pe\~narrubia et al. (2008) 
 become unstable on time scales of a few 10$^8$ yr. 
Thus we set up a new suite of simulations (Burkert et al. in prep.), in which we can reproduce the appearance of HCC-087 in an environment as the Hydra cluster. 
\subsubsection{HCC-087 orbiting HCC-005}
In practice, we model the dwarf as a one-component\footnote{Simulations with a concentrated stellar component embedded in an extended dark matter halo reach identical conclusions, i.e., 
the stars trace the disruption of the dark halo and outline the observed, characteristic S-shape. A prerequisite is an {\em a priori} large extent of the stellar component.} 
Hernquist (1990) profile with $r=0.5$ kpc and $1\times10^7$ M$_{\odot}$ (Misgeld \& Hilker 2011)\footnote{
We also ran simulations with Plummer profiles and verified that the results do not depend on the shape of the profile as long as the satellite is strongly tidally perturbed during peri-center passage.} 
and place it in an 
orbit around a massive, HCC-005-like  galaxy with M$_{\rm tot}=5\times10^{11}$ M$_{\odot}$ (ibd.)\footnote{
Note that this is the total mass, accounting for the galaxy's dark halo. This value is consistent with the observed stellar mass of $3.4\times10^{11}$ M$_{\odot}$
(Misgeld \& Hilker 2011), assuming a dark matter-to-baryon fraction similar to the MW.}. 

Surprisingly, our simulations indicate that an S-shape of the disrupting satellite, just as observed for HCC-087,  naturally emerges 
at the {\em first apocenter} (Fig.~4), which, in the simulations, occurs at 50 kpc -- coinciding with the (projected) present-day distance between HCC-087 and 
HCC-005.
More strikingly, the orientation of the tidal tails at apocenter is {\em perpendicular} to the orbit.
\begin{figure}[tb]
\begin{center}
\includegraphics[angle=90,width=1.5\hsize]{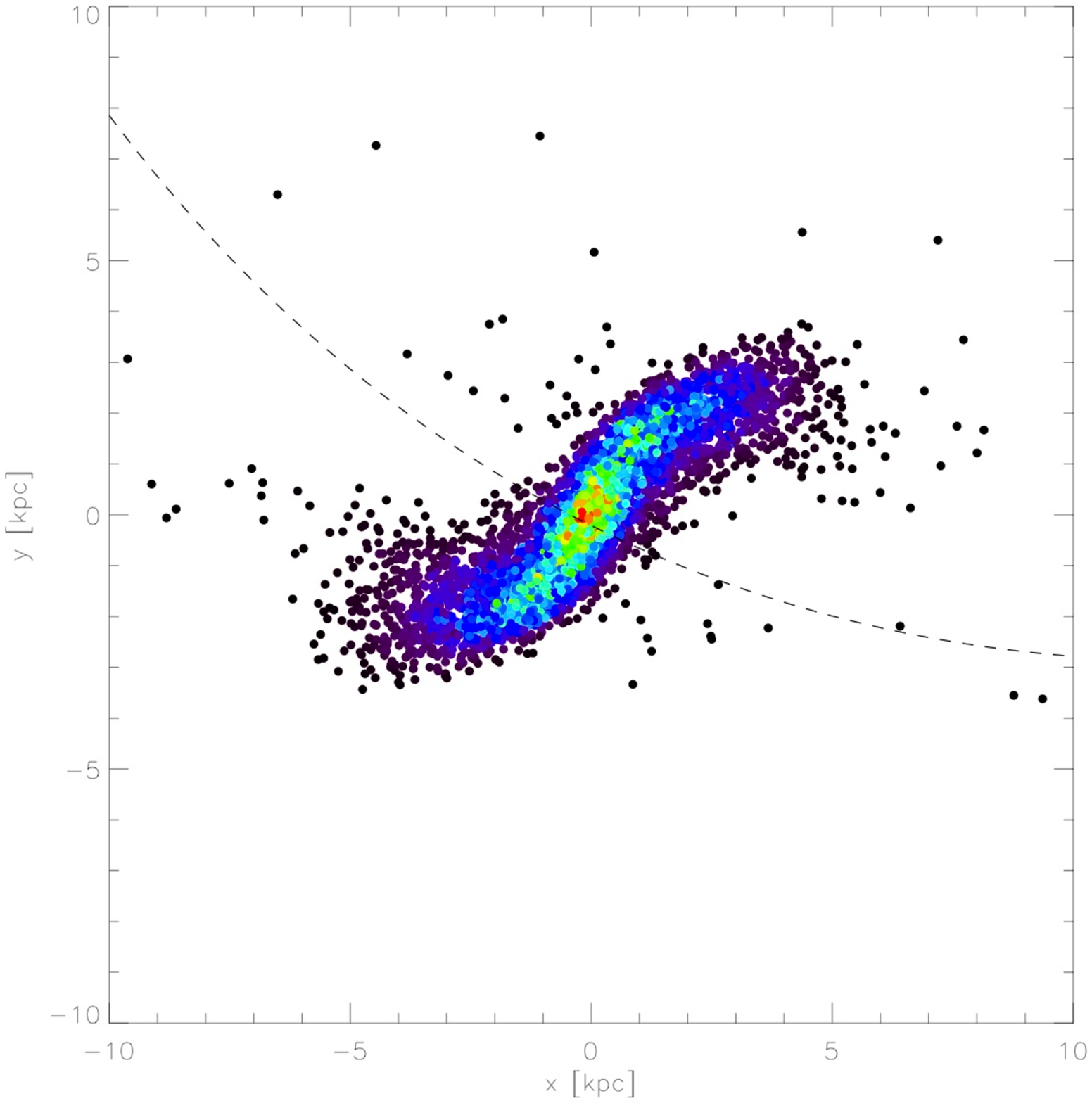}
\includegraphics[angle=90,width=1.5\hsize]{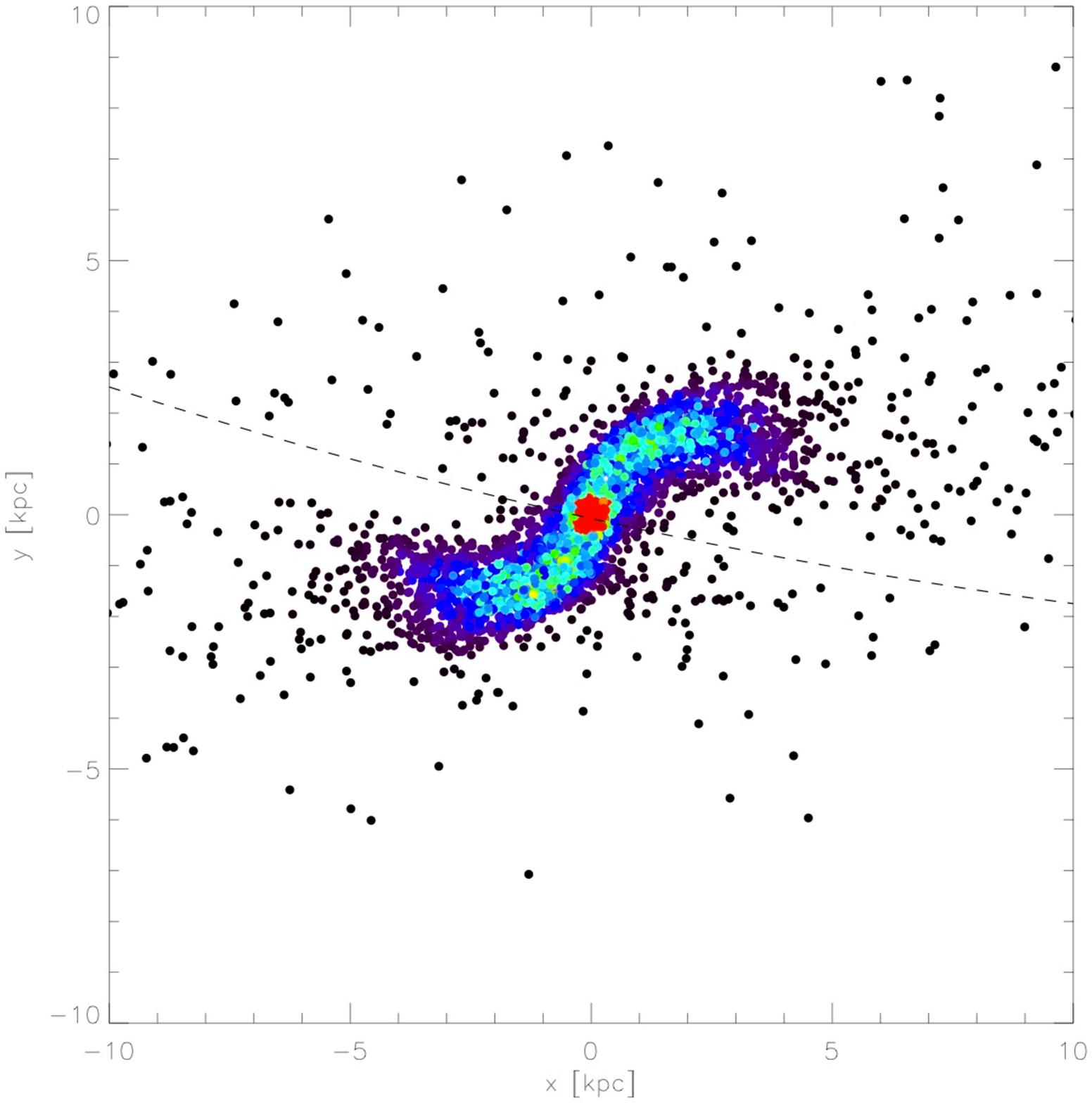}
\end{center}
\vspace{-0.4cm}
\caption{Snapshot of the simulation of HCC-087 orbiting the Hydra I galaxy HCC-005 (top panel) and the Hydra cluster center (bottom panel), 
color-coded by stellar density. The observed morphology is reproduced  at first apocenter,  each after $\sim0.8$ Gyr.}
\end{figure}

Its having past the first apocenter passage at 50 kpc then begs the question of the origin of HCC-087. We note that ``first-infall'' events are not uncommon in the Local Group 
as several high-velocity dwarf galaxies make their passages through the M31 or MW halos (Chapman et al. 2007; Majewski et al. 2007; Lepine et al. 2011). 
None of these, however, displays any evidence of tidal tails. 
Therefore, we investigated the possibility that the observed apocenter   could be a result of dynamical friction, adopting a radius and mass-to-light ratio 
as for MW dSphs of comparable luminosities (Koch et al. 2007). 
We find that dynamical friction is able to bring HCC-087 on a bound orbit as observed after one orbital period, while starting at the 
virial radius of the host galaxy. This requires, however, a very small relative velocity of the interactors ($\sim 100$ km\,s$^{-1}$), which is in contrast to the high virial velocity 
of the relaxed cluster ($\sim 750$ km\,s$^{-1}$; M08). 
\subsubsection{HCC-087 orbiting the Hydra I potential}
As an alternative, we placed HCC-087 on an eccentric  and a circular orbit around the cluster center itself, represented by the location of the primary galaxy NGC~3311 
(e.g., Arnaboldi et al. 2012). 
To this end, we 
computed the total baryonic mass of Hydra~I within 150 kpc (the present position of HCC-087), using the data of Christlein \& Zabludoff, as 2$\times10^{12}$ M$_{\odot}$ and 
a dark matter mass 10--25 times as large (we note that the results turn out insensitive to the exact value).  
This set-up reproduces Hydra's 
virial velocity at 750 km\,s$^{-1}$ and places HCC-087 at a starting distance of 150 kpc with a pericenter of 50 kpc  in the eccentric case (Burkert et al., in prep.). 

As a result, this scenario is also able to reproduce the morphology, orientation,  and spatial extent of HCC-087, 
in full accord with Fig.~4. However, contrary to the interaction with HCC-005 (Sect.~3.3.1), 
a disruption by the cluster itself is bolstered by the higher relative velocity of the dSph and the parent, which is more realistic given the high velocity dispersion of the cluster. 
Nonetheless, the comparison with the 
set-up in Sect.~3.3.1 and the similar case of NGC~4449B indicates that the morphology of the object is largely independent of the host of the disruption. 
Likewise, the shape and density profile of the satellite  at first apocenter  
is close to indistinguishable between    the cases of an orbit around HCC-005 and the cluster center  (Fig.~4). 
\section{Summary}
We report on the re-analysis of deep imaging data of the Hydra I galaxy cluster, on which we detect the clear signatures of tidal disruption 
on one of the HCC early-type dwarfs. Its large spatial extent of 3.1 kpc renders it one of the most extended dwarf galaxies in the Local Volume, 
in line with NGC 4449B, And~XIX,  and our in-house, prime example, Sagittarius.  

HCC-005 lies roughly on  a line connecting HCC-087 and the cluster center -- a constellation that will inevitably put HCC-087's morphological evolution into a predicament. 
Overall, we emphasize the significance of its disruption in that the cluster center exerts such a devastating effect on the satellite despite its large distance. 

Aided by a new suite of N-body simulations we argue that the dwarf's morphology is caused by 
orbiting  the galaxy cluster center. 
An orientation of the tails perpendicular to the orbit is naturally reproduced at the 
current apocenter at 150 kpc after an interaction time scale of $\sim$0.6 Gyr  on an eccentric orbit, and similarly for several Gyr when we assume an almost circular orbit. 

While, morphologically,  we cannot rule out that HCC-087 was disrupted by the closer (50 kpc) early-type galaxy HCC-005, 
the very low relative velocity (less than 15\% the cluster's virial velocity)
 required for 
such an interaction renders this scenario unlikely. 
Nonetheless, our simulations have quantitatively shown that the  structure of the satellite alone is no unique indicator of the system it is orbiting. 
\acknowledgments
AK thanks the Deutsche Forschungsgemeinschaft for funding from  Emmy-Noether grant  Ko 4161/1. 
\end{document}